\documentclass[aps,showpacs]{revtex4}
\usepackage[dvips]{graphicx}
\usepackage{amsmath}
\newcommand{\be}{\begin{equation}}
\newcommand{\ee}{\end{equation}}
\newcommand{\ba}{\begin{align}}
\newcommand{\ea}{\end{align}}
\newcommand{\bitem}{\begin{itemize}}
\newcommand{\eitem}{\end{itemize}}

\newcommand{\benum}{\begin{enumerate}}
\newcommand{\eenum}{\end{enumerate}}
\newcommand{\bc}{\begin{center}}
\newcommand{\ec}{\end{center}}

\begin{document}

\title{HOW SCALING SYMMETRY SOLVES A SECOND-ORDER DIFFERENTIAL EQUATION}
\author{Sidney Bludman}
\email{sbludman@yahoo.com}
\homepage{http://www.das.uchile.cl/~sbludman}
\affiliation{Departamento de Astronom\'ia, Universidad de Chile, Santiago, Chile}
\author{Andres Guzman}
\email{aguzman@das.uchile.cl}
\affiliation{Departamento de Astronom\'ia, Universidad de Chile, Santiago, Chile}
\author{Dallas C. Kennedy}
\email{dalet@stanfordalumni.org}
\homepage{http://home.earthlink.net/~dckennedy}
\noaffiliation
\date{\today}

\begin{abstract}
While not generally a conservation law, any symmetry of the equations of
motion implies a useful reduction of any second-order equationto a first-order equation between invariants, whose solutions (first
integrals) can then be integrated by quadrature (Lie's Theorem on the
solvability of differential equations). We illustrate this theorem by
applying scale invariance to the equations for the hydrostatic equilibrium
of stars in local thermodynamic equilibrium: Scaling symmetry reduces the
Lane-Emden equation to a first-order equation between scale invariants
$u_n, v_n$, whose phase diagram encapsulate all the properties of index-n
polytropes.  From this reduced equation, we obtain the regular (Emden)
solutions and demonstrate graphically how they transform under scale
transformations.
\end{abstract}


\maketitle
\tableofcontents
\section{SYMMETRY REDUCES THE ORDER OF ANY SECOND ORDER ODE} 

Lie showed how the invariance of a second-order ordinary differential
equation (ODE) under a point symmetry leads constructively to a first-order
ODE plus a quadrature.  If the symmetry were a variational symmetry of the
Action Principle, Noether's well-known theorem would lead to a conservation
law \cite{Olver2,Boyce,JordonSmith}.

We consider the \emph{scaling symmetry} $\xi \rightarrow A\xi,~\theta_n
(\xi) \rightarrow \theta_{nA} (\xi)$ of the second-order ODE
\emph{Lane-Emden equation} (Section III) which describes the hydrostatic
equilibrium of a gaseous sphere or star in local thermodynamic equilibrium
(Section II). Because this is not a variational symmetry, but only a
symmetry of the Lane-Emden equation, scaling symmetry leads only to a
\emph{non-conservation law } \cite{BludKenI,BludKenII}.  This is a
first-order differential equation for scale invariants $u_n,~v_n$, which
can be solved for $v_n (u_n)$ for given boundary conditions (Section III).
From these first integrals, quadrature finally leads to solutions of the
original second-order equation (Section IV).

Sections II and III will
consider only \emph{regular solutions} which have density finite at the
origin and apply only to complete polytropes.
Section IV will generalize to \emph{irregular solutions} which have densities
infinite at the origin (F-solutions) or vanish away from the origin
(M-solutions) and can apply only to stellar envelopes.

\section{HYDROSTATIC EQUILIBRIUM OF SELF-GRAVITATING SPHERES} 

An adiabatic sphere in hydrostatic equilibrium obeys the equations of
equilibrium between gravitational and internal (pressure)forces and of mass
continuity \be -d P/\rho dr= G m/r^2 , \quad d m/d r=4\pi r^2 \rho \quad,
\label{eq:masspressbalance} \ee where the local pressure, mass density, and
included mass $P(r),~\rho(r),~m(r)$ depend on radius $r$.  In terms of the
the gravitational potential $V(r)=\int_\infty^r Gm/r^2 dr$ and thermostatic
potential (specific enthalpy, ejection energy) $H(r)=\int_{0}^{P(r)}
dP/\rho $, (\ref{eq:masspressbalance}) and its integrated form \be -d H/d
r=d V/d r ,\quad V(r)+H(r)=-\frac{GM}{R}\quad, \ee

\begin{figure}[t] 
\includegraphics[scale=0.65]{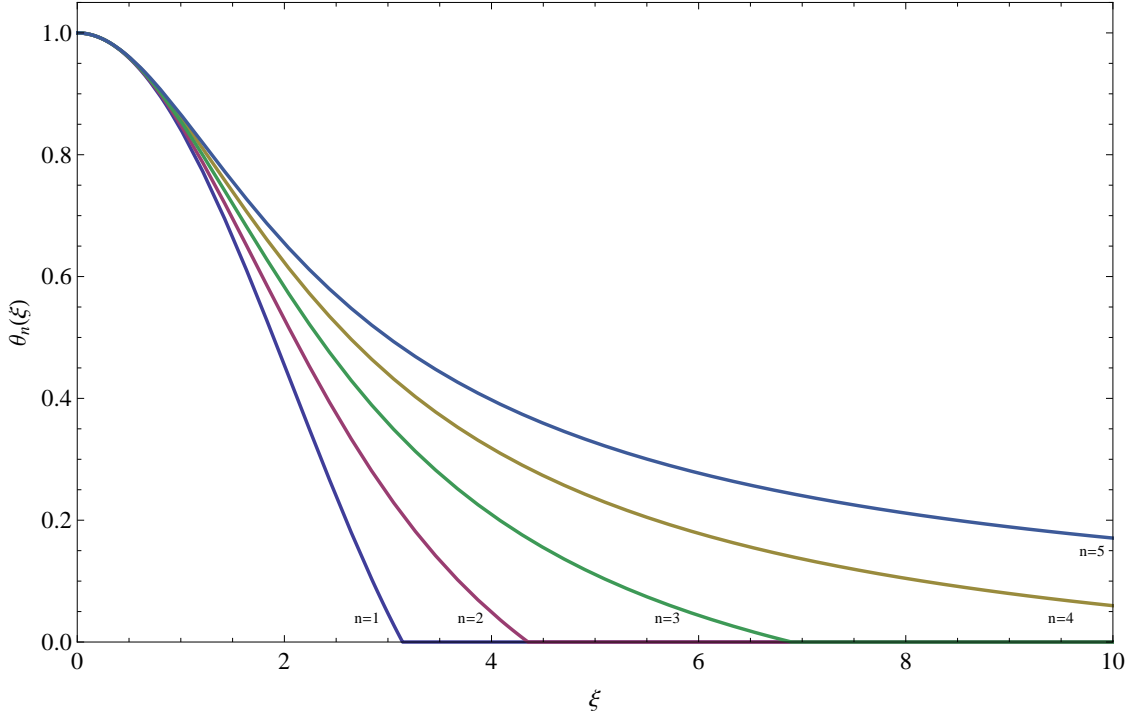} 
\caption{Emden Functions (Regular Solutions of the Lane-Emden Equation) of
order n=1,2,3,4,5}
\end{figure}

\begin{figure}[t] 
\includegraphics[scale=0.65]{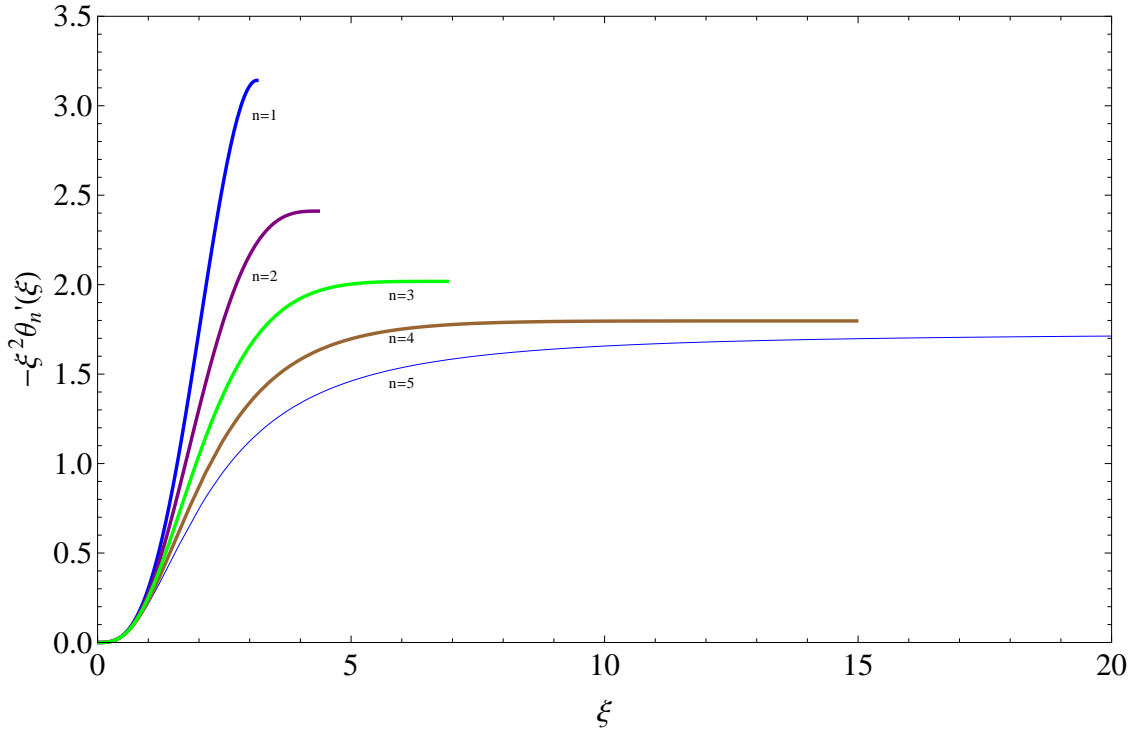}
\caption{'Mass' included inside radius $\xi$ for complete polytropes of
order n=1,2,3,4,5 terminating at $\xi_{1n}=3.141,4.353,
6.897,14.97,\infty$.}
\end{figure}

expresses the conservation of the specific energy as sum of gravitational
and internal energies, in a star of mass $M$ and radius $R$. The two
first-order equations (\ref{eq:masspressbalance}) can be combined into the
second-order Poisson's Law \be \frac{1}{r^2}\frac{d}{dr} \Big( r^2\frac{d
H}{dr} \Big) + 4 \pi G \rho(H)=0 \quad , \label{eq:secondorder} \ee 
in terms of the enthalpy $H(r)$.

The two equations (\ref{eq:masspressbalance}) can always be written
\be
d\log{u}/d\log{r}=3-u(r)-n(r) v(r)\quad ,\quad %
 d\log{w}/d\log{r}=u-1+v(r)-d\log{[1+n(r)]}/d\log{r}\quad, \label{eq:equil}
\ee
in terms of the logarithmic derivatives
\be
u(r):= d\log{m}/d\log{r} \\
w(r):= n(r) v(r)= -d\log{\rho}/d\log{r} \\
v(r):= -d\log{(P/\rho)}/d\log{r}
\ee 
and an index $n(r)$ \be n(r):=d\log{\rho}/d \log{(P/\rho)} \quad,\quad
1+1/n(r):=d\log{P} /d\log{\rho}\quad ,\ee which depends on the local adiabatic
equation of state $P=P(\rho)$.

For solutions regular at the origin, spherical symmetry requires that $dP/dr=0$ and that $\rho(r),~P(r),~H(r)$ be even functions of $r$. Mass
continuity requires, to order $r^2$, \be \rho(r)\approx\rho_c (1-Ar^2),
~~m(r)\approx\frac{4 \pi r^3}{3}\cdot(1-\frac{3}{5}A r^2)\approx\frac{4 \pi
r^3}{3} \cdot \rho_c^{2/5} \rho(r) ^{3/5} \quad . \label{eq:origin} \ee
Thus, near the origin, the average mass density inside radius
$\bar{\rho}(r):=\frac{m(r)}{4\pi r^3/3}\approx\rho_c^{2/5} \rho(r) ^{3/5}$
and $u(0)=3, v(0)=0$. At the stellar surface, $u(R)=0, v(R)=\infty$.
Here
$\bar{\rho}(r):=\frac{m(r)}{4\pi r^3/3}$ is the average mass density inside radius, so that
$u(r)=3\rho/\bar{\rho}$ decreases from $u(0)=3$ at the origin to $u(R)=\infty$ at the stellar surface. The ratio
$v(r):=\frac{3}{2}(-Gm/r)/(\text{P} / \rho )=\frac{3}{2}
(\text{'gravitational energy'})/(\text{'internal energy of ideal gas'})$ increases from $v(0)=0$ at the origin to $v(R)=\infty$ at the stellar surface.

\section{SCALE TRANSFORMATIONS ON                                                                                                                                                        SOLUTIONS OF THE LANE-EMDEN EQUATION} 

The symmetry we consider is scaling symmetry, the most general
simplification that one can make for any dynamical system.  If a scale
transformation $r\rightarrow Ar$ transforms $m,~\rho, ~P$ multiplicatively,
the logarithmic derivatives $u,~v$ will be scale invariant.  The
structural equations (5) will then be autonomous, if and only if
$n=constant$, so that, $P(r)=K\rho(r)^{1+\frac{1}{n}}$, with the constant
$K$ determined by the constant specific entropy.  When this is so, scaling
symmetry leads to the two coupled autonomous equations for the scale
invariants \be du_n/d\log{r}=u_n (3-u_n-n v_n)\quad ,\quad
dv_n/d\log{r}=v_n(u_n-1+v_n)\quad, \label{eq:chareqns12} \ee 
implying the first-order equation
\be
dv_n/du_n=v_n(u_n-1+v_n)/u_n (3-u_n-n v_n)\quad. \label{eq:chareqns13} \ee
This reduced equation for $v_n(u_n)$ incorporates all the consequences of scaling
symmetry, from which all the properties of polytropes follow.

In this and the next section, we consider only \emph{regular solutions}
(Emden or E-solutions) of the Lane-Emden equation, which have finite
density at the origin, and are applicable only to complete polytropes.
At the origin, the initial conditions on regular solutions of \ref{eq:chareqns13} are
$u_n (0)=3, v_n (0)=0$. (We defer to Section IV the irregular solutions $v_n (u_n)$, where the density at the origin is not finite.)

Figure 3 shows how, moving radially outwards, $u_n(r)=3
\rho(r)/\bar{\rho(r)}$ decreases from $3$ at the origin to $0$ at the outer
boundary $R$ and $v_n(r)$ increases from $0$ to $\infty$. The combinations $\omega_n(r):=(u v_n^n)^{1/(n-1)}\equiv
-\xi^{1+\tilde{\omega}_n} \theta_n '$ approach the finite values
$_0\omega_n (R)$ and characterize each n-polytrope.

\begin{figure}[t]  
\includegraphics[scale=0.60]{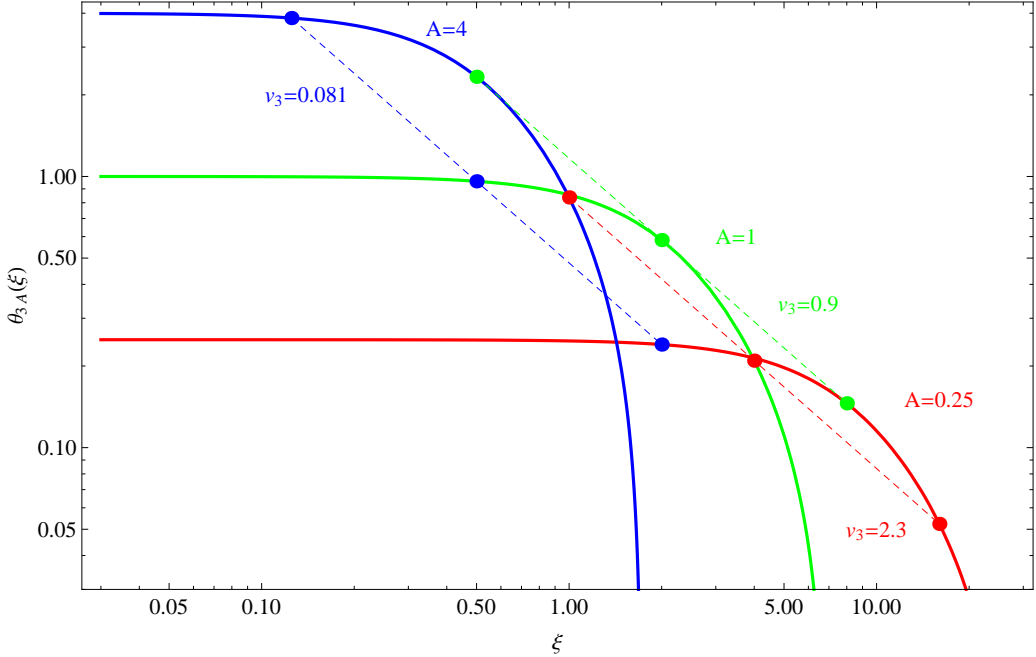}
\caption{Emden function $\theta_3(\xi,A)$ for three scales A=4 (blue),
1 (green), .25 (red).  On a log-log plot, rescaling simply transforms the
Lane-Emden functions along lines of constant
$v_3:=-d\log{(P/\rho)}/d\log{r}$. The three dashed lines connect homologous
points for the three illustrative values $v_3=0.081, 0.25, 2.3$.}
\end{figure}

In terms of the dimensional central density, pressure $\rho_c,~P_c=K \rho_c ^{1+1/n}
$ and constant $\alpha ^2:=((n+1)/{4\pi G})K\rho_c ^{1/n-1},
~H_c:=(n+1) (P/\rho)_{c} \equiv (n+1) K \rho^{1/n}_c $, the dimensional
radius, enthalpy, mass density and included mass are \be
r=\alpha \xi,\quad H=H_c\theta_n, \quad \rho=\rho_c \theta_n ^n,\quad
m(r)=(4 \pi \rho_c \alpha ^3 )(-\xi^2 \theta_n ') \quad ,\ee where prime
designates the derivative $':=d/d \xi$. The scale invariants are
\be u_n:=-\xi {\theta_n}^n/\theta_n' , \qquad v_n := -\xi\theta_n '/\theta_n  \qquad \omega_n:=(uv^n)^{1/(n-1)}=-\xi^{\frac{n=1}{n-1}}\quad,\ee
where $\tilde{\omega_n}:=2/(n-1)$.

In these dimensionless units,                                ,
Poisson's Law (\ref{eq:secondorder}) becomes the {\emph {Lane-Emden
equation}} \cite{Chandra} \be \frac{d}{d\xi}\Bigl( \xi^2
\frac{d\theta_n}{d\xi}\Bigr) + \xi^2\theta_n ^n = 0
\label{eq:lane-emden-1}, \ee whose regular solutions
have
\be \theta_n=const, \qquad \theta_n ' (0) =0,\qquad u(0)=3,\qquad v_n(0) =0 \quad , \ee
The \emph{normalized regular solutions} with $\theta_n=1$
define the {\em Emden Functions} for the n-polytrope radial evolution
(Figure 1). Their dimensionless normalized 'mass' $-\theta^2 \theta_n '
(\xi) $ is shown in Figure 2. The first order equations (\ref{eq:chareqns12}) have analytic solutions
listed in Table 1 for $n=0, 1, 5$, but must be integrated numerically for any
other polytropic index $n$.
\begin{figure}[t] 
\includegraphics[scale=0.60]{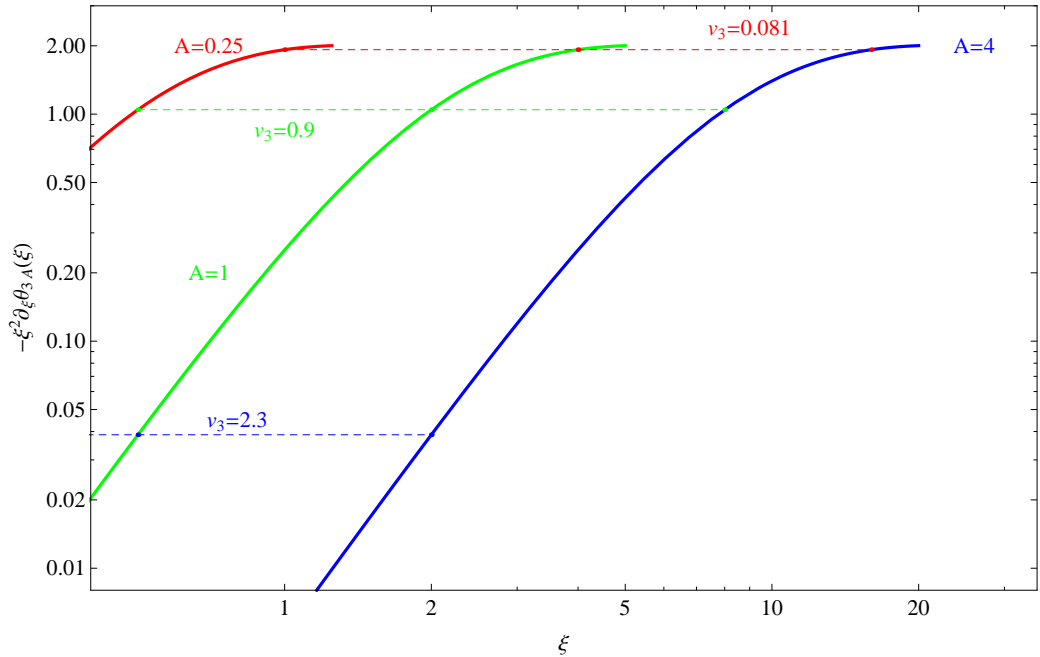}
\caption{'Mass' included inside radius $\xi$ for n=3 polytrope. On a
log-log plot, rescaling simply transforms the 'mass' along lines of
constant $v_3:=-d\log{(P/\rho)}/d\log{r}$. The dashed lines connect
homologous points for the three illustrative values $v_3=0.081, 0.25, 2.3$.}
\end{figure}

The Lane-Emden equation (\ref{eq:lane-emden-1}) is invariant under the
scale transformation $\xi\rightarrow A\xi,~t\rightarrow t+\log{A},
~\theta_n(\xi)\rightarrow A^{\tilde{\omega}_n} \theta_n(A~\xi):=\theta_{n
A} (\xi) $. Besides the Emden Function $\theta_n(\xi)$, normalized so that
$\theta_{n1} (\xi)\equiv\theta_n (\xi)$, the Lane-Emden equation has
rescaled regular solutions $\theta_{nA} (\xi)$, whose value at the origin
is $A^{\tilde{\omega_n}}$. Figures 3 and 4 show log-log plots of the
rescaled n=3 function $\theta_{3A}(\xi)$ and rescaled 'mass' $-\theta'^2
\vartheta_\xi \theta_{3A} (\xi)$ for three different rescalings A=4 (blue), 1
(green), 0.25 (red). On a log-log plot, rescaling appears as a translation
along the dashed lines of constant $v_n$ and $u_n$, which connect homologous
points. All the familiar properties of polytropes \cite{Chandra} follow from this scaling
symmetry incorporated in the reduced equation for $v_n(u_n)$.

\begin{table}[h]   
\caption {\textbf{Scaling Invariants and normalized Emden Functions for n=0, 1, 5}}
\begin{tabular}{|l||l||l|l||l|l||l|r|}
\hline\hline
$n$                        &$v_n(u_n)$                        &$u_n(\xi)$                   &$v_n(\xi)$          &$\theta_n(\xi)$      &$-\xi^2 \theta_n'(\xi)=\xi^{\frac{(n+1)}{(n-1)}}\omega_n$   &$\xi_{1n}$                &$ _0 \omega_n$  \\

\hline\hline
0                          &$u=3$                             &3                            &$2\xi^2/(6-\xi^2)$  &$1-\xi^2/6$          &$\xi^3/3$                                              &2.45                      &0.333          \\
1                          &parametric $u_n(\xi),v_n(\xi)$    &$\xi^2/(1-\xi\cot{\xi})$     &$1-\xi\cot{\xi}$    &$\sin{\xi}/\xi$      &$\sin{\xi}-\xi\cos{\xi}$
&3.14                      &$\cdots$       \\
5                          &$1-u_5/3$                         &$3/(1+\xi^2 /3)$               &$\xi^2 /(3+\xi^2)$&$(1+\xi^2 /3)^{-1/2}$&$\xi^3/3(1+\xi^2 /3)^{3/2}$
&$\infty$                  &0              \\
\hline\hline
\end{tabular}
\end{table}

\begin{figure}[t] 
\includegraphics[scale=0.60]{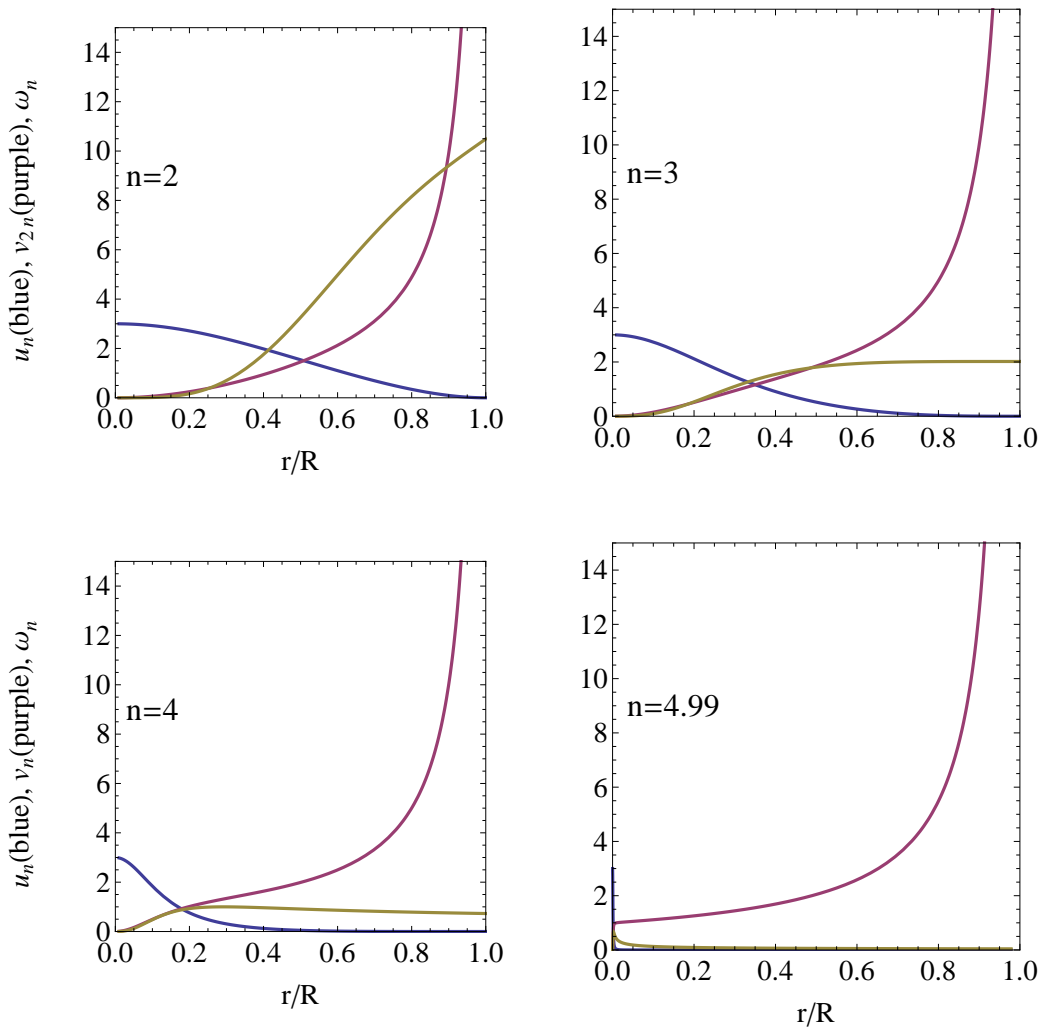}
\caption{The Emden scale invariants $u_n(r)=d\log{m}/d\log{r}, v_n(r)=-d\log{P}/d\log{r}$ and
$\omega_n (r)=(uv^n)^{1/(n-1)}$. While $v_n(r)$ diverges at the stellar radius $R$, $\omega_n$ (in green) approaches the finite values $_0 \omega_2=10.49,  ~_0 \omega_3=2.02,
~_0 \omega_4=0.73,
~_0 \omega_{4.99} \approx 0$ characteristic of each n-polytrope.
The curves in the $n=4.99$ figure are excellent approximations to the $n=5$ functions on the bottom line of Table I for which $R=\infty$.}
\end{figure}

\section{PHASE DIAGRAM FOR ALL FIRST INTEGRALS} 

Equation (\ref{eq:chareqns13}) and the integral curves $v_n (u_n)$ and $u_n (v_n)$ are scale invariant functions of the scale invariant (r/R) plotted in
Figure 5 for n=2,~3,~4,~4.99.
All the dependence on scale $R$ is contained in (\ref{eq:chareqns12})which reads \be \log{(r/R)}=\int^\infty_0
\frac{dv_n} {v_n(u_n(v_n)-1+v_n)}=\int^0_3 \frac{du_n} {u_n(3-u_n-n v_n
(u_n))} \quad \ee
for the \emph{regular solutions} obeying $u_n(0)=3, v_n(0)=0$ at the
origin.

The \emph{irregular solutions} of (\ref{eq:chareqns13}) apply only to incomplete
polytropes, the scale-invariant envelopes of stars.  As shown in Figure 6, these
are \emph{F-solutions} (green curves, $u_n (0)>3$) which have infinite density at the
origin and \emph{M-solutions }(blue curves, $v_n (0)$) which have density vanishing
before the origin ($v(0)>0$). These irregular solutions would obtain by
integrating in from the boundary values $_F \omega_n > _0 \omega_n$ and  $_0 \omega_n> _M
\omega_n $. Their separatrices (red curves) are the regular (Emden) solutions, which have finite density
at the origin, $u_n(0)=3,~v_n(0)=0$ and boundary values $_0 \omega_n $.

\section{CONCLUDING SUMMARY}

A symmetry of the equations of
motion generally implies, not a conservation law, but a still
useful reduction of order of the equations of motion to a
first-order equation between invariants, which can then be integrated by
quadrature.

For the Lane-Emden equation, the reduced equation is the first-order
equation (\ref{eq:chareqns13}), whose first integrals (Figure 6) encapsulate
all the properties of index-n polytropes.  From this reduced equation
(Figure 5), we obtained the regular (Emden) solutions (Figures 1, 2) and
simply demonstrated their scale dependence on log-log plots (Figures 3, 4).

\begin{figure}[t] 
\includegraphics[scale=0.60]{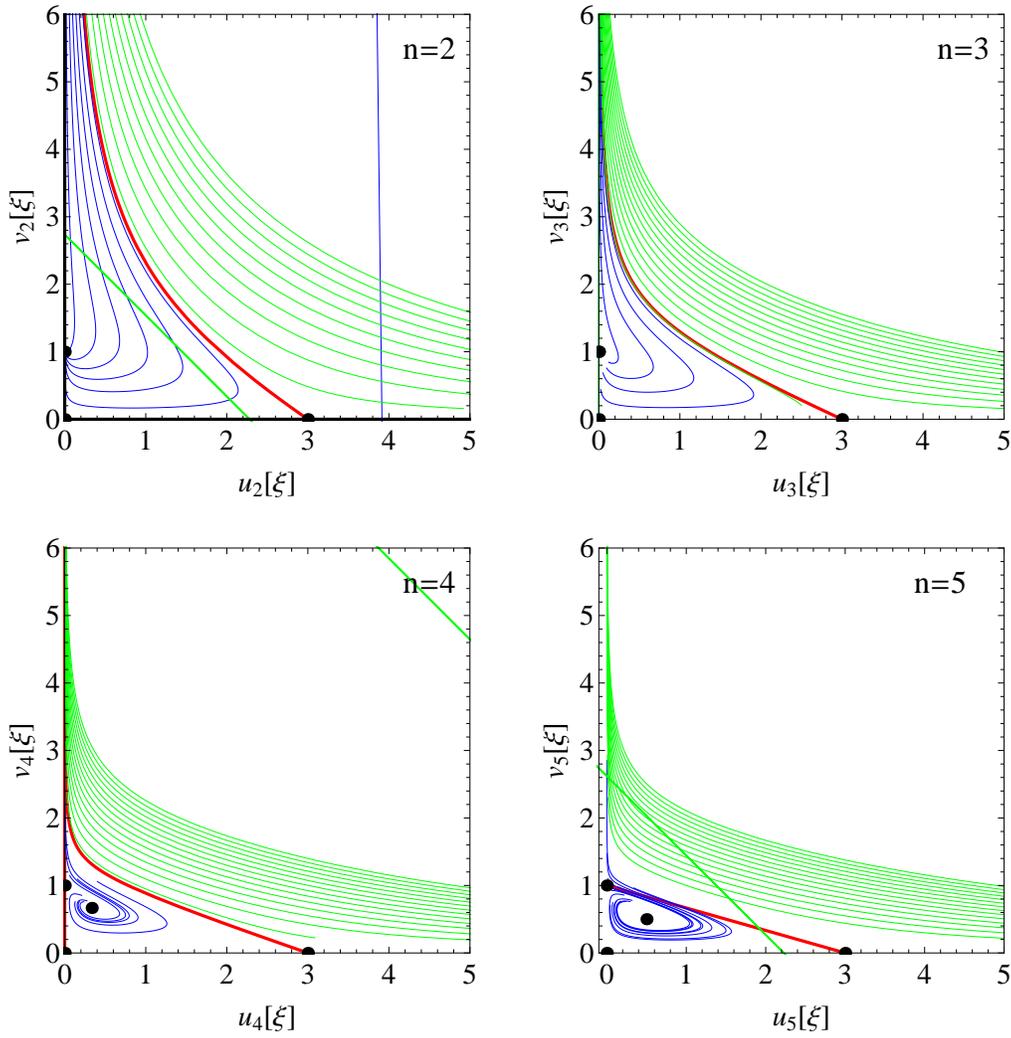}
\caption{Critical Points and Phase Diagrams for $v_n(u_n)$ for n=2, 3, 4, 5.  The green curves
(F-solutions) have infinite density at the stellar center.  The blue curves (M-solutions) have
vanishing density away from the origin.  Their separatrix, the red curve (Emden solutions), has finite
density at the origin. }
\end{figure}

\bibliography{bibliographyLE}

\end{document}